\documentclass{article}
\usepackage{icmc2026_paper_template}
\usepackage{times}
\usepackage{ifpdf}
\usepackage{soul}
\usepackage{amsmath}
\usepackage{placeins}
\usepackage[english]{babel}
\usepackage{comment}

\def\papertitle{Perpetual Dialogues: A Computational Analysis of Voice–Guitar Interaction in Carlos Paredes's Discography}
\def\firstauthor{Gilberto Bernardes}
\def\secondauthor{Nádia Moura}
\def\thirdauthor{António Sá Pinto}

\newif\ifpdf
\ifx\pdfoutput\relax
\else
   \ifcase\pdfoutput
      \pdffalse
   \else
      \pdftrue
  \fi
\fi

\ifpdf 
  \usepackage[pdftex,
    pdftitle={\papertitle},
    pdfauthor={\firstauthor, \secondauthor, \thirdauthor},
    bookmarksnumbered, 
    pdfstartview=XYZ 
   ]{hyperref}

  \usepackage[pdftex]{graphicx}
  \graphicspath{{./figures/}}
  \DeclareGraphicsExtensions{.pdf,.jpeg,.png}

  \usepackage[figure,table]{hypcap}

\else 
  \usepackage[dvips,
    bookmarksnumbered, 
    pdfstartview=XYZ 
  ]{hyperref}  

  \usepackage[dvips]{epsfig,graphicx}
  \graphicspath{{./figures/}}
  \DeclareGraphicsExtensions{.eps}

  \usepackage[figure,table]{hypcap}
\fi

\hypersetup{
    colorlinks,%
    citecolor=black,%
    filecolor=black,%
    linkcolor=black,%
    urlcolor=black
}

\title{\papertitle}

\threeauthors
  {0.5in}
  {\firstauthor} {INESC TEC, Faculty of\\ Engineering, University of Porto\\ 
     {\tt \href{mailto:gba@fe.up.pt}{gba@fe.up.pt}}}
  {\secondauthor} {CEIS20, Faculty of Arts and\\ Humanities, University of Coimbra \\ %
     {\tt \href{mailto:nmoura@fl.uc.pt}{nmoura@fl.uc.pt}}}
  {\thirdauthor} {INESC TEC, Faculty of\\Engineering, University of Porto\\ %
     {\tt \href{mailto:asapinto@fe.up.pt}{asapinto@fe.up.pt}}}

\begin{document}
\sloppy
\capstartfalse
\maketitle
\capstarttrue
\begin{abstract}
Computational musicology enables systematic analysis of performative and structural traits in recorded music, yet existing approaches remain largely tailored to notated, score‑based repertoires. This study advances a methodology for analyzing voice–guitar interaction in Carlos Paredes’s vocal collaborations—an oral‑tradition context where compositional and performative layers co‑emerge. Using source‑separated stems, physics‑informed harmonic modelling, and beat‑level audio descriptors, we examine melodic, harmonic, and rhythmic relationships across eight recordings with four singers. Our commonality–diversity framework, combining multi‑scale correlation analysis with residual‑based detection of structural deviations, reveals that expressive coordination is predominantly piece‑specific rather than corpus‑wide. Diversity events systematically align with formal boundaries and textural shifts, demonstrating that the proposed approach can identify musically salient reorganizations with minimal human annotation. The framework further offers a generalizable computational strategy for repertoires without notated blueprints, extending Music Performance Analysis into oral-tradition and improvisation-inflected practices.

\end{abstract}
\section{Introduction}\label{sec:introduction}

In ``blueprint''-driven performances—those relying on documented structures such as scores—interpretative decisions primarily involve nuanced changes in timing and dynamics while leaving compositional content intact. Music Performance Analysis (MPA)\footnote{While MPA extends beyond the analysis of musical audio and encompasses multiple dimensions such as visuals and staging, this study focuses on audio analysis.} has traditionally focused on such contexts, comparing multiple performances of a shared blueprint to study expressive deviation and interpretative choice. This approach has naturally favored Western art music, where notated scores and extensive performance corpora enable systematic comparative analysis~\cite{palmer1997music,MouraVidal2023}. 


Conversely, MPA faces significant challenges when addressing repertoires rooted in oral transmission, popular music, or improvisation, where no explicit blueprint exists, and compositional and performative dimensions are inseparable~\cite{prouty2006orality,tani2024traditional}. 
In such practices, musical ideas emerge, evolve, and are shaped through performance itself, with recordings often capturing relatively stable realizations of these processes rather than interpretations of a pre-existing score~\cite{rosenblum2023varying}.

Within this context, the work of Carlos Paredes provides a particularly revealing case. Widely recognized as a solo composer–performer, Paredes transformed the Portuguese guitar from a primarily accompaniment role into an autonomous musical voice, redefining the expressive potential of Portuguese musical traditions in the twentieth century. This aesthetic position emerged from a formative trajectory grounded in collective practice: born into a lineage of Coimbra guitarists, Paredes learned the instrument through oral transmission from his father, Artur Paredes, and began his career as a second guitarist in ensemble contexts, where he already demonstrated a marked tendency to reshape and elaborate accompaniments beyond conventional models \cite{Correia2016CancaoCoimbra, Fonseca2025CarlosParedes}. The resulting tension between accompaniment and artistic autonomy underpins both his solo output and his vocal collaborations.

Our work establishes the first steps toward the study of Carlos Paredes’s recorded vocal collaborations, some framed in the Portuguese \emph{Canção de Coimbra} tradition, spanning his entire career. Although these collaborations represent only a small fraction of his largely solo discography, they provide a consistent and well-defined setting in which voice and Portuguese guitar form two stable interacting layers. We adopt audio content analysis to address both \emph{performative} and \emph{structural} dimensions, which—despite their somewhat artificial separation—help frame the agentic interplay through which compositional and interpretative processes co‑shape these works.

By leveraging audio source separation, we isolate these two layers to enable independent analysis of their melodic, harmonic, and rhythmic characteristics. This approach allows us to explore how these layers interact across different works, revealing patterns of dialogue, contrast, and complementarity that define Paredes's collaborative style. Our analytical framework centers on statistical tools that capture commonalities—through correlations and linear relationships—and diversity—through residual analysis and outlier detection.

Given this perspective, our research addresses the following questions: (i) What performative and structural characteristics distinguish individual pieces, specific collaborations, and the corpus as a whole? (ii) What patterns of voice–guitar interaction—through commonality and diversity analysis—can be identified across Paredes’s collaborative recordings, and how do these patterns vary between recordings and collaborators?

This work contributes to MPA in three key ways. First, it introduces an integrated framework that jointly models structural and performative dimensions in repertoires where no notated blueprint exists. Second, the approach combines source separation, physics‑informed harmonic descriptors, and multi‑scale statistical analysis to reveal interaction patterns that are otherwise obscured in polyphonic recordings. Third, by framing commonality and diversity through both correlations and residual‑based deviations, the method identifies not only stable expressive relationships but also structurally meaningful departures that align with perceptually salient musical events. 

The remainder of this paper is structured as follows. Section~\ref{Sec:rel_work} reviews relevant work in MPA, source separation, and structural analysis. Section~\ref{sec:corpus} introduces the corpus of Paredes’s vocal collaborations. Section~\ref{sec:comp_methods} presents the computational pipeline, including source separation, descriptor extraction, and statistical methods. Section~\ref{sec:descriptor_analysis} reports the assessment of our descriptors. Section~\ref{sec:results} details the main analytical findings on melodic, harmonic, and rhythmic characteristics and their interactions. Section~\ref{sec:discussion} provides a musicological interpretation of the results considering the broader context of Paredes’s artistic practice. Section~\ref{sec:conclusions} concludes with implications for the computational study of oral‑tradition–based repertoires and outlines directions for future work.

\section{Related Work}
\label{Sec:rel_work}

MPA has traditionally focused on score-based traditions where interpretative decisions involve nuanced changes in timing and dynamics while preserving notated compositional content~\cite{cancinochacon2018comprehensive,palmer1997music}. Computational models of expressive performance enable systematic evaluation of performance hypotheses through quantitative analysis~\cite{cancinochacon2018comprehensive}. However, these approaches are fundamentally biased toward repertoires with explicit notational blueprints, making them less applicable to musical traditions rooted in oral transmission or improvisational practices~\cite{prouty2006orality,tani2024traditional}.

In traditions where composition and performance are inseparable, such as jazz, popular music, and other orally transmitted practices, the continuum from strictly notated to fully improvised music reflects distinct music traditions and social 
dynamics ~\cite{rosenblum2023varying, christodoulou2025amd}. In oral musical cultures, improvisation involves the internal process of recalling previously heard music rather than manipulating pre-existing notated material~\cite{tani2024traditional}. This distinction is particularly relevant for analyzing Carlos Paredes's work within the Canção de Coimbra tradition, where compositional ideas emerge and crystallize through performance and recording rather than through predetermined scores~\cite{Fonseca2025CarlosParedes, Correia2016CancaoCoimbra}.

Computational musicology has made substantial progress in automated music analysis through audio content descriptors and machine learning approaches~\cite{meredith2016computational}. Music source separation—the task of decomposing recordings into constituent instrumental or vocal components—has seen tremendous advancement with deep learning methods~\cite{chen2024indepth}. These techniques divide into model-based approaches that explicitly represent musical structure through repetition or harmonic patterns~\cite{schulzeforster2022unsupervised}, and data-driven deep learning methods that achieve state-of-the-art performance given sufficient training data~\cite{defossez2021hybrid,stoter2019openunmix}. Recent work has applied deep learning frameworks to specific vocal ensemble challenges, including the analysis of choral singing and multi-voice separation~\cite{chandna2022deeplearning}.

Following source separation, structural analysis techniques identify musical organization through repetition detection and pattern discovery~\cite{dannenberg2002pattern,paulus2010music}. The automatic extraction of repeating patterns enables music summarization, indexing, and structural segmentation~\cite{lu2004repeating,muller2015music}. These methods typically rely on self-similarity matrices computed from acoustic descriptors to detect recurring segments at multiple temporal scales.
However, most computational approaches focus on identifying structural commonality through repetition, with less emphasis on explicitly modeling diversity or performative dimensions beyond timing variations~\cite{flexer2012novelty}.

The integration of structural and performative analysis remains underexplored, particularly for repertoires lacking explicit scores. While some studies examine rhythm-pitch interactions in source-separated tracks~\cite{dittmar2012audio}, performance aspects such as dynamics, articulation, and expressive phrasing receive limited attention in computational frameworks. This gap likely reflects both the predominance of structure-focused research questions and the scarcity of datasets with detailed performance annotations~\cite{cancinochacon2018comprehensive,christodoulou2024multimodal}.

Our study addresses these limitations by applying integrated structural and performative analysis to Carlos Paredes's vocal collaborations—a corpus representing the intersection of composed and improvised elements within the Portuguese guitar tradition. By leveraging audio source separation to isolate vocal and guitar layers, we enable independent melodic, harmonic, and rhythmic characterization while examining their interaction patterns through statistical analysis of correlations and residuals. This approach extends MPA methodology to oral-tradition-based popular music, where the artificial distinction between structural and performative dimensions remains essential for computational tractability while acknowledging their fundamental inseparability in practice.

\section{Paredes's Collaborations Corpus} 
\label{sec:corpus}

While Carlos Paredes’s discography is predominantly centered on solo guitar works, it includes six collaborations with singers between 1958 and 1983 (a total of 26 tracks, as documented in the complete discography published in \cite{Paredes2002Integral}). Precisely because of their limited number, these recordings constitute a well-defined and analytically relevant corpus for examining guitar–voice interaction in the \textit{Canção de Coimbra} tradition. The present study intentionally adopts a narrow subset of the corpus, focusing on eight pieces (two songs per collaboration) to establish the methodological framework. The following recordings were selected:
\begin{itemize}
    \item Augusto Camacho, \textit{Fado de Coimbra} (1958): `A Água da Fonte', `Adeus a Coimbra', `A Luz do Teu Olhar', and `Quando os Sinos Dobram';
    \item Luiz Goes, \textit{Coimbra de Ontem e de Hoje} (1967): `Balada do Mar' and `Canção da Infância';
    \item Cecília de Melo, \textit{Meu País} (1970): `Não Choro por me Deixares' and `Render dos Heróis';
    \item Adriano Correia de Oliveira, \textit{Que Nunca Mais} (1975): `Recado a Helena' and `Tejo que Levas as Águas'. 
\end{itemize}

This selection features three singers from Canção de Coimbra's tradition and one lyrical singer (Melo), whose stylistic background contrasts with the popular and fado-related traditions represented in the rest of the corpus. This focused set serves to demonstrate the pipeline that will be used to analyze the entire collection of collaborations.

The corpus was annotated using Sonic Visualiser \cite{SonicVisualiser}. Annotations were initially carried out independently by two musicologists to address subjective bias. A subsequent review stage involving a third musicologist was used to consolidate discrepancies and compile a shared annotation set, which served as the basis for the analyses presented in this study. Four annotation layers were defined: (1) beat and bar-level timing; (2) formal structure, capturing large-scale sectional organization; (3) tonal phrase segmentation, describing melodic contour units; and (4) harmonic progression, including chordal changes and harmonic functions.

\section{Computational methods}\label{sec:comp_methods}
\begin{figure}
    \centering
    \includegraphics[width=1\linewidth]{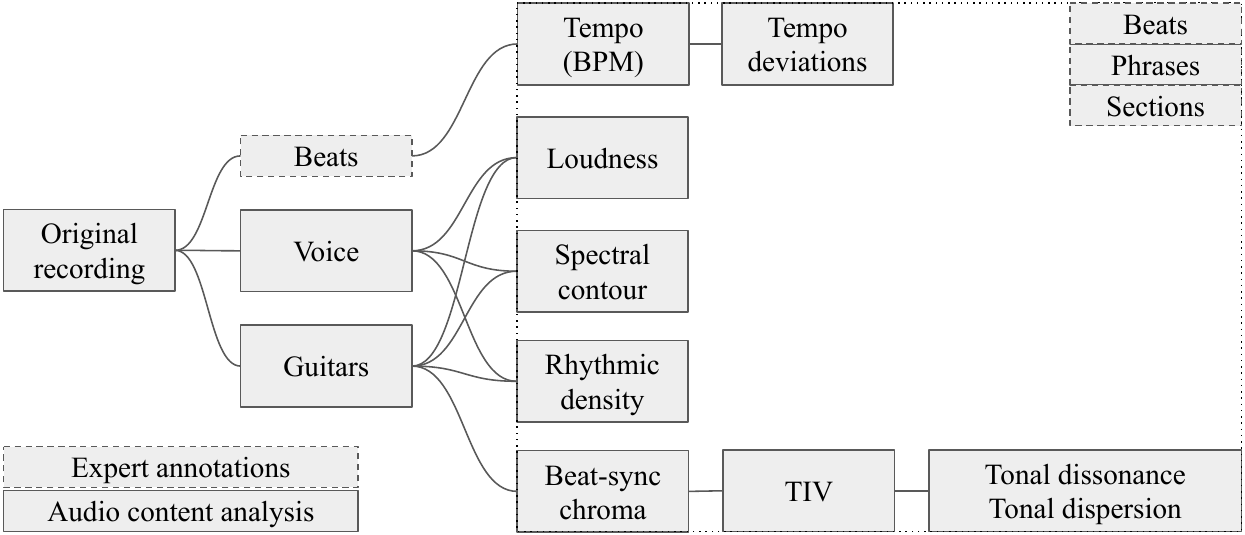}
    \caption{Computational pipeline for the extraction of musical descriptors. Dashed boxes indicate expert-driven manual annotations used to compute specific descriptors and to impose hierarchical segmentations.}
    \label{fig:diagram}
\end{figure}

Figure~\ref{fig:diagram} presents the computational pipeline employed in this work. Our input data comprise the original track (uncompressed audio at 44.1 kHz sample rate) and expert temporal segmentation annotations for beats, phrases, and sections. Prior to audio content analysis, we apply the Ht-Demucs model~\cite{rouard2023hybrid} to separate the vocal and guitar tracks into two distinct audio streams.
The guitar track is further decomposed into harmonic and percussive components using the harmonic-percussive source separation (HPSS) algorithm by Fitzgerald~\cite{fitzgerald2010harmonic}. We refer to these separated audio streams as `guitar harmonic' and `guitar percussive' components, respectively. Our implementation is available online at: \url{https://github.com/bernardes7/Paredes}. Unless otherwise specified, the framework relies on version 0.11 of the \texttt{librosa} library~\cite{mcfee2015librosa} for audio processing and \texttt{ScyPy}~\cite{2020SciPy-NMeth} library for statistical and time series processing.  

\subsection{Audio Content Analysis}

For each dimension under analysis, we extract a time series at beat-level resolution. This resolution was chosen as a suitable compromise between human interpretability and computational efficiency, given that the genre under study exhibits limited nuanced expression below the beat level.
 
\textbf{Dynamic descriptors} capture the temporal evolution of loudness across the voice and guitar tracks separately, computed as the root mean square energy per frame and converted to decibels (dB) relative to the maximum amplitude to ensure comparability across sources and enhance perceptual representation. These frame-level descriptors are then aggregated at beat positions. Finally, a loudness mask for voice and guitars is computed to identify regions below -40 dB, denoting inactive segments, which are then used to filter out all remaining descriptors.

\textbf{Rhythmic descriptors} include the tempo curve, overall tempo, tempo deviations, and rhythmic density (computed per voice and guitar parts). The tempo curve is obtained by converting the expert inter-beat intervals annotations into beats per minute (BPM). From this curve, we estimate the piece’s tempo as the median value within the first section where voice and guitars interact (typically, the second section following an introduction by the guitars). This section is assumed to represent the baseline tempo established by the performers during their initial interaction. 
Rhythmic density refers to the number of note onsets per beat. For the guitar part, it is computed from a flux-driven novelty detection applied to the guitar percussive component. For the voice part, it is obtained through note transcription based on the pYIN algorithm, followed by post-processing using Hidden Markov Models (HMM), as implemented in the corresponding Vamp plugin \cite{mauch2014pyin}.

\textbf{Melodic and harmonic contour descriptors} are derived from the voice and harmonic guitar tracks, respectively. We compute these contours as the spectral centroid of the Constant-Q Transform (CQT) representation of each track, providing a systematic inference under the assumption of relatively stable timbre. While this approach mitigates inference errors potentially introduced by adopting advanced pitch-tracking algorithms, only the relative curve shape can be interpreted and not the absolute pitch height, as the centroid reflects spectral energy distribution rather than precise fundamental frequency. Inactive beats in both melodic and harmonic curves are excluded by applying the respective dynamic masks.

Remaining \textbf{harmonic descriptors}, computed from the guitar’s harmonic track, include dissonance and tonal dispersion.  The baseline representation is a frame chroma, derived from the guitar’s harmonic track, obtained through a pipeline incorporating CQT analysis, denoising, whitening, and Non-Negative Least Squares (NNLS) decomposition using an instrument-specific dictionary tailored to the specificities of the Portuguese and fado guitars found in literature~\cite{inacio2004vibroacoustic,debut2016physics,serraosilva2017experimental}. This process requires parameterizing the NNLS dictionary to reflect the complex physical and acoustic properties inherent to these instruments.

The dictionary incorporates key physical acoustic behaviors. Harmonic amplitudes $a_k$ decrease exponentially with harmonic order, controlled by a decay rate factor $s$~\cite{debut2016physics}. Odd harmonics are selectively boosted by a factor $\gamma > 1$ to capture the bright metallic timbre characteristic of these guitars~\cite{debut2016physics}. Inharmonicity is introduced to account for string stiffness, causing partial frequencies $f_k$ to deviate slightly from integer multiples of the fundamental $f_0$ by a small coefficient $\beta$~\cite{debut2016physics}. Furthermore, body resonances are modeled as Gaussian gains $g(f)$ centered at specific frequencies $f_r$ (at 120, 470, and 1000 Hz, with $\sigma$ widths of 30, 80 and 150 Hz, respectively) to reflect the substantial soundboard and air cavity amplification~\cite{debut2016physics,serraosilva2017experimental}. $S_{res}$ represents the resonance gain scaling factor\footnote{
The lowest center, 120 Hz, approximates the highly efficient air cavity mode (Helmholtz resonance) experimentally identified around $125 \text{ Hz}$ in Coimbra models~\cite{serraosilva2017experimental}. The intermediate center, 470 Hz, is crucial for encompassing the major structural resonance group observed in the 250-500 Hz range, specifically covering higher longitudinal dipole modes (e.g., at 406 and 458 Hz) that contribute significantly to the radiated sound spectrum. The highest frequency center, 1 kHz, models the substantial high-frequency energy contributions, utilizing a broad gain function to capture the brilliance and clarity required for the distinct metallic timbre of the instruments.}. The complete formulation for the final partial amplitude $A_k$ is:
\begin{equation}
\begin{aligned}
a_k &= s^{(k-1)},\\
f_k &= f_0 \cdot k \cdot \sqrt{1 + \beta k^2},\\
a_k' &=
\begin{cases}
a_k \cdot \gamma, & \text{$k$ odd},\\
a_k, & \text{$k$ even},
\end{cases}\\
g(f) &= S_{res} \cdot \sum_{r=1}^{R} G_r \cdot \exp\!\Big(-\dfrac{(f - f_r)^2}{2\sigma_r^2}\Big),\\
A_k &= a_k' \cdot g(f_k).
\end{aligned}
\end{equation}

After median beat aggregation, Tonal Interval Vectors (TIV) are computed as the as the discrete Fourier transform (DFT) of beat-synchronized chromas, whose coefficients are weighted by empirical consonance ratings~\cite{bernardes2016multi}. This transformation enables higher-level harmonic analysis grounded in perceptual principles and aligns with music theory frameworks using the DFT~\cite{amiot2016music}. From these TIVs, we derive tonal dissonance and tonal dispersion using the \texttt{tiv.lib} library~\cite{ramires2020tiv, bernardes2017automatic}. To define a harmonic center, from which the tonal dispersion Euclidean distances are computed, we adopt a TIV vector resulting from the centroid of all beat-synchronous TIV of the section where guitars and voice first interact.

\subsection{Statistical Analysis}
\label{sec:stats}

We quantify commonalities and diversity in performance by analysing pairwise linear dependencies among the extracted musical descriptors. Because descriptor time series are typically non-stationary and contain multi-scale fluctuations, we first apply empirical mode decomposition (EMD) and focus on slower-varying components by reconstructing each descriptor from IMFs 3 and higher, $x^{(\ge 3)}(t)=\sum_{j\ge 3}\mathrm{IMF}_j(t)$. This reduces the influence of rapid local fluctuations and stabilises dependence estimates at the macro-structural level.

For each song, we compute the Pearson correlation matrix across descriptors using the reconstructed signals $x^{(\ge 3)}(t)$. To obtain collaboration- and corpus-level summaries, correlations are aggregated across songs using Fisher’s z-transform: $z_k=\operatorname{atanh}(r_k), \bar z=\frac{1}{K}\sum_k z_k, \text{and} \, \bar r=\tanh(\bar z)$, preserving sign and avoiding the bias induced by direct averaging of $r$. Descriptor pairs are retained as candidate interaction patterns when the aggregated correlation exceeds a practical effect-size threshold (e.g., $|\bar r|>0.6)$; in supplementary analyses we verify robustness using bootstrap confidence intervals\footnote{We evaluated whether the cutoff $|r|\geq 0.6$ was appropriate by estimating 95\% bootstrap confidence intervals (moving‑block bootstrap, $B=1000$) for every pairwise correlation across all pieces. No feature pair in the corpus exhibited a confidence interval entirely above 0.6, whereas a substantial majority had confidence intervals entirely below this threshold. The remaining near‑threshold cases displayed wide confidence intervals crossing 0.6 and highly variable bootstrap probabilities of exceeding the threshold, indicating that these correlations are not statistically stable. Thus, 0.6 acts as a conservative but empirically justified boundary: correlations below it are reliably weak, while correlations above it would require evidence stronger than what is present in the data.}.

To identify moments of diversity relative to an established dependency, we model the within-song relationship between each retained descriptor pair by robust linear regression, $y(t)=a+bx(t)+\varepsilon(t)$, and examine the time-resolved residuals $\varepsilon(t)$. Residuals are standardized using a robust scale estimate (median absolute deviation), and extreme deviations are flagged when $|z(t)|>2.5$. Positive residual outliers indicate transient exaggeration beyond the typical linear coupling (values above the fitted relation), whereas negative residual outliers indicate transient attenuation or reversal (values below the fitted relation).

\begin{figure}
    \centering
    \includegraphics[width=1\linewidth]{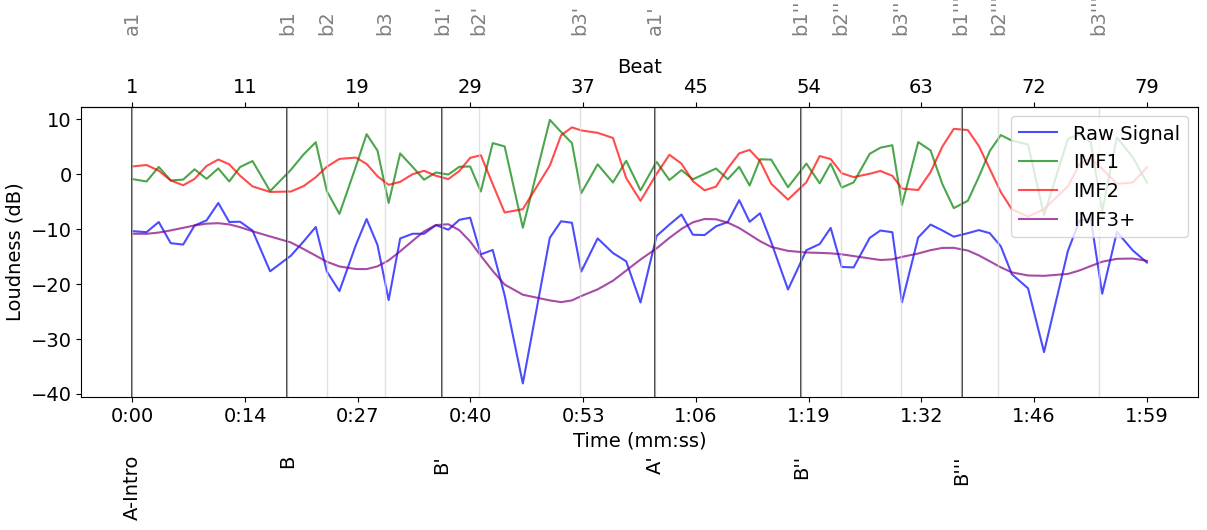}
    \caption{Empirical Mode Decomposition of the guitar loudness descriptor in `Balada do Mar' by Paredes and Goes, showing the raw signal alongside intrinsic mode functions IMF 1, IMF 2, and IMF 3 and higher.}
    \label{fig:placeholder}
\end{figure}

\section{Descriptors assessment}
\label{sec:descriptor_analysis}

Given that all signal-processing descriptors derive from separated stems, we conducted a qualitative assessment of the Ht-Demucs outputs to ensure that artefacts did not systematically bias analyses. Separated tracks were reviewed by two expert listeners to identify separation failures. Although occasional artefacts were observed, they were predominantly confined to high‑energy consonants or rapid ornamental figures, providing sufficient grounds for interpreting descriptor‑level correlations as reflections of musical interaction.

The descriptors adopted in this work are well-documented and evaluated in related literature. However, a significant refinement was introduced in the harmonic descriptors. The quality of the beat-synchronous chroma representation was enhanced by incorporating a specialized dictionary specifically adapted to the Portuguese and fado guitars. To validate this adaptation and quantify the efficacy of the physical modeling, a preliminary evaluation was conducted via an ablation study. 

This study measured the contribution of various physics-informed components to the NNLS beat-synchronous chroma estimation accuracy, benchmarked against binary chroma representations derived from expert annotations\footnote{A perfect alignment with the binary chroma is not expected due to the presence of additional spectral components inherent to instrumental performance, including overtones, transients, and noise-related partials.}. The ablation analysis systematically investigated the contribution of six harmonic modeling components: frequency whitening (WHI) and median filtering (MED); and dictionary factors including harmonic decay (DEC), $s$, odd-harmonic boost (ODD), $\gamma$, body resonances (RES), and inharmonicity (INH) ($\beta$). Following individual component assessment, Bayesian optimization with Gaussian process surrogates fine-tuned three progressive configurations using cosine similarity: (A) $\text{WHI+DEC}$; (B) $\text{WHI+DEC+ODD+RES}$; and (C) the full model ($\text{WHI+DEC+ODD+RES+INH+MED}$).

The analysis showed that WHI and DEC were the most impactful individual components and their combination (A) achieved mean cosine similarity of .739. Progressive optimization revealed that the highest score of .741 was attained by (B), slightly outperforming (C), the full model scoring .738. This suggests the addition of INH and MED slightly diminished performance. Optimal parameters for (B) included zero spectral detuning, maximum decay strength ($s =.5$), a reduced odd harmonic boost of $\gamma = .95$, and maximized resonance gain scale of $1.4$. These results underscore the critical importance of modeling strong harmonic decay and utilizing prominent body resonance frequencies ($\text{120.0 Hz}$, $\text{470.0 Hz}$, $\text{1000.0 Hz}$) to capture the instrument's distinct characteristics.

\section{Results}\label{sec:results}

\begin{figure}[h]
\centering
\includegraphics[width=0.86\columnwidth]{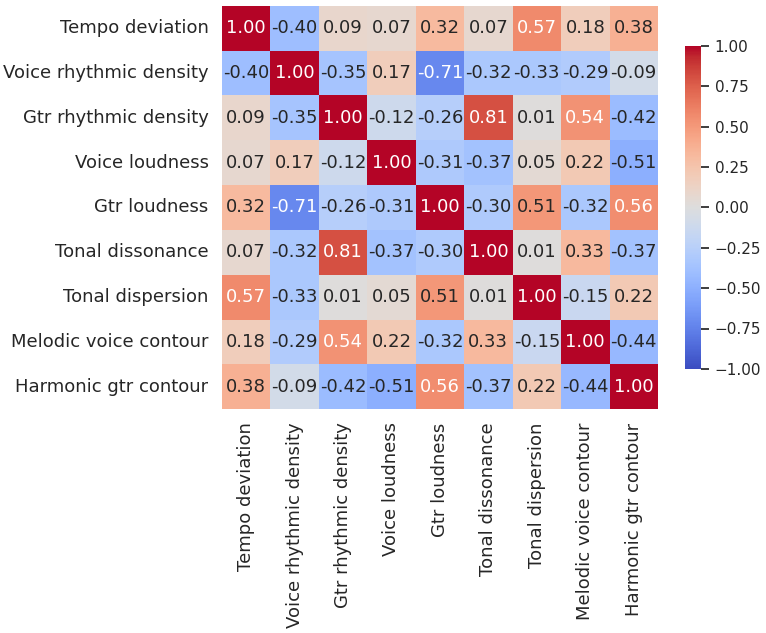}
\caption{Correlation matrix for the song ‘Balada do Mar’ (Paredes and Goes), aggregated using Fisher’s z-transformation of Pearson correlations.\label{fig:corr_goesSong}}
\end{figure}

While the complete set of results is available online in the complementary materials to this paper\footnote{The correlation matrix per piece is made available at \url{https://github.com/bernardes7/Paredes}.}, we focus here on the interpretation of correlations in `Balada do Mar', a collaboration between Paredes and Goes, shown in Figure~\ref{fig:corr_goesSong}, as an illustrative example of our analytical workflow. 

\begin{figure}[h]
    \centering
    \includegraphics[width=1\linewidth]{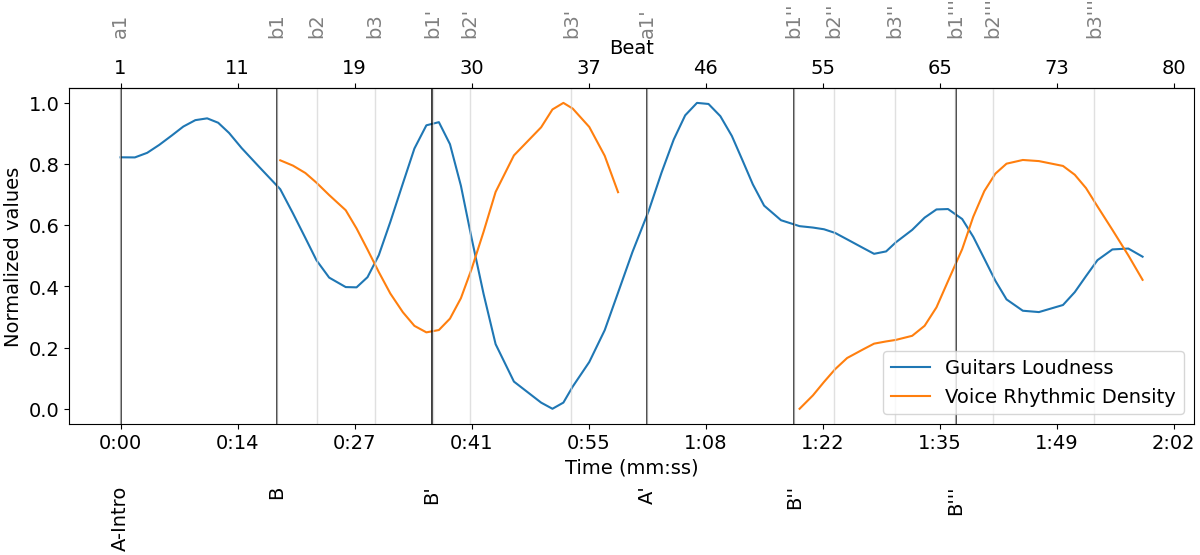}
    \caption{Time series of the Empirical Mode Decomposition (IMF 3+) for guitar loudness and voice rhythmic density in `Balada do Mar' by Paredes and Goes. Masked segments correspond to periods of vocal inactivity below -40 dB. Vertical black lines denote section boundaries, while vertical grey lines denote phrase boundaries. Section labels are displayed at the bottom of the plot, and phrase labels are displayed at the top.}
    \label{fig:time_seriesGoes}
\end{figure}

\begin{figure}
    \centering
    \includegraphics[width=1\linewidth]{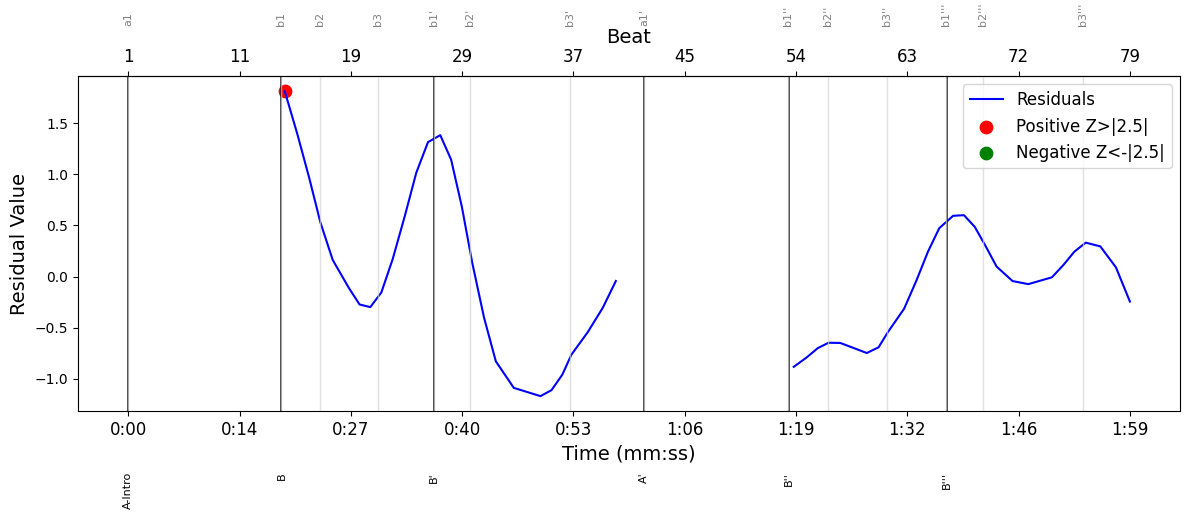}
    \caption{Residual analysis of the guitar loudness and voice rhythmic density interaction in `Balada do Mar' by Paredes and Goes. The highlighted point identifies a positive outlier with a residual z-score exceeding $+2.5$, indicating a breakdown of the typical contrapuntal relationship at the beginning of section B. Vertical black lines denote section boundaries, while vertical grey lines denote phrase boundaries. Section labels are displayed at the bottom of the plot, and phrase labels are displayed at the top.}
    \label{fig:residual}
\end{figure}

Two descriptor relationships of particular interest are the correlation between tonal dissonance and guitar rhythmic density ($r=.81$) and between guitar loudness and voice rhythmic density ($r=-.71$). Figure~\ref{fig:time_seriesGoes} shows these two time series overlaid to facilitate inspection of their correlation in musical time. As indicated by the negative correlation, the articulation of melodic onsets in the voice exhibits contrapuntal motion (mirrored behavior) with the presence of the guitars, as measured by their loudness. An exception to this correlation is observed at the beginning of section B, where both parameters exhibit parallel motion from high to low density and loudness, as shown in the positive outlier indicated in the residual analysis (see Figure~\ref{fig:residual}). 

Once the relationships and divergent points of interest were identified, a perceptual inspection of these patterns was conducted to validate the computational findings. This analysis confirmed that the dense guitar ostinato typically fills the space during sustained or silent vocal passages, with an exception at the beginning of section B, where voice and guitar articulate simultaneously.


\begin{figure*}
    \centering
    \includegraphics[width=1\linewidth]{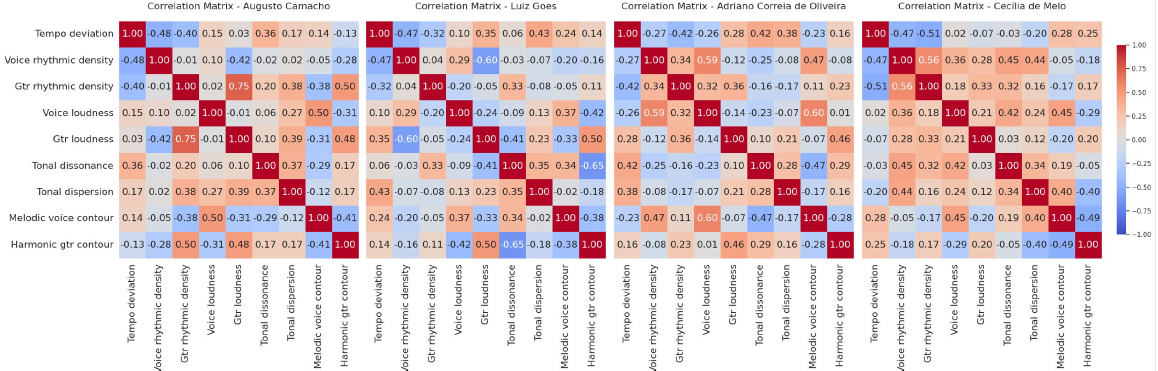}
    \caption{Correlation matrices for each Paredes vocal collaboration, aggregated using Fisher's z-transformation of Pearson correlations.}
    \label{fig:global_colab_mtx}
\end{figure*}

\begin{figure}
    \centering
    \includegraphics[width=0.86\linewidth]{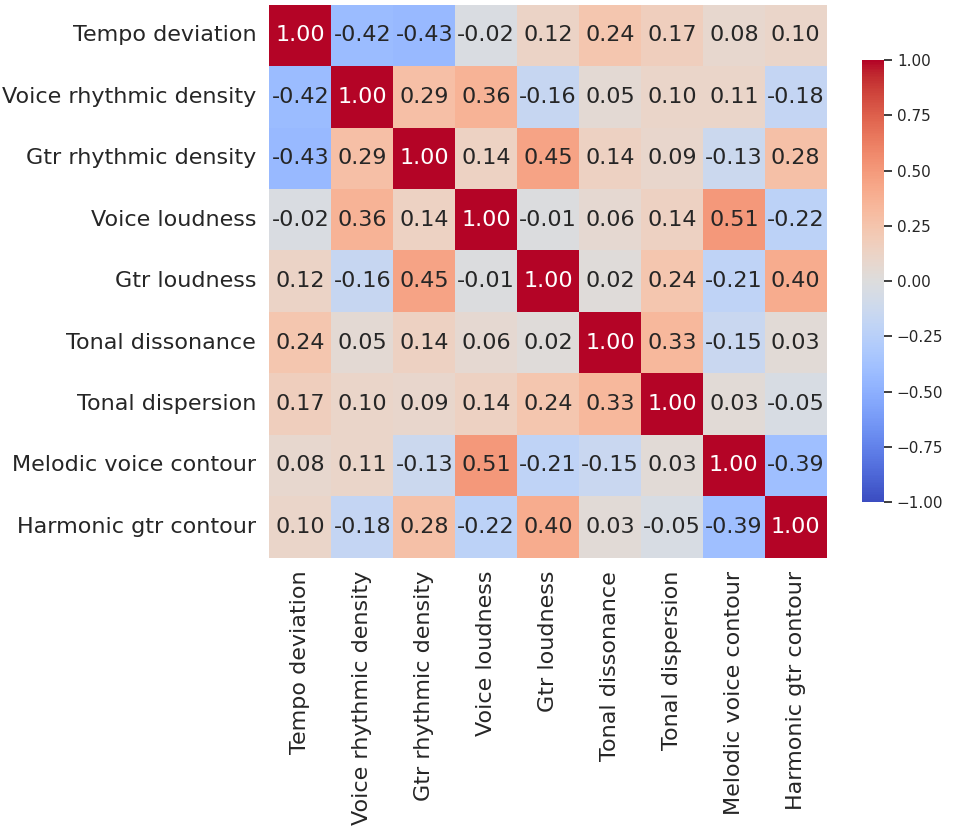}
    \caption{Correlation matrix for all Paredes vocal collaboration adopted in our study, aggregated using Fisher's z-transformation of Pearson correlations.}
    \label{fig:global_mtx}
\end{figure}

Figure~\ref{fig:global_colab_mtx} presents the aggregated matrices per collaboration using Fisher’s z-transform. Examining the matrices at the level of individual collaborations reveals a set of interactional signatures that are both internally consistent and context-dependent. In the collaboration with Camacho, the strongest correlations involve the guitar layer (guitar rhythmic density-guitar loudness, $r=.75$; guitar rhythmic density-harmonic guitar contour, $r=.5$), alongside the moderate correlation between voice rhythmical density and voice loudness ($r=.5$). The Goes collaboration shows correlations involving cross-layer and harmonic descriptors: between voice rhythmic density and guitar loudness ($r=-.6$), tonal dissonance and harmonic guitar contour ($r=-.65$), and guitar loudness and harmonic guitar contour ($r=.5$). In collaboration with Correia de Oliveira, expressive correlations are restricted to the vocal layer (voice rhythmic density-voice loudness, $r=.59$; voice melodic contour-voice loudness, $r=.6$). The Melo collaboration shows a negative correlation between guitar rhythmic density and tempo deviation ($r=-.51$) and a positive correlation between guitar rhythmic density and voice rhythmic density ($r=.56$). Taken together, these results show that the strongest linear dependencies differ across collaborations both in magnitude and in the descriptor domain.

When aggregating all songs across collaborations into a single global correlation matrix (Figure~\ref{fig:global_mtx}), no dominant interaction patterns emerge (correlations remain within the interval $[-.43,\,.51]$). The absence of strong, consistent correlations at this level suggests that there is no single, corpus-wide marker governing Paredes’s vocal collaborations. Instead, interaction strategies appear to be adapted to the specific combination of singer, repertoire, and piece-level musical structure. 

The strongest correlation that remains perceptible at the global level concerns the relationship between vocal melodic contour and vocal loudness ($r=.51$), where higher pitch regions tend to coincide with increased intensity. Although this pattern reflects a general tendency for vocal performance rather than a guitar–voice dependency, its persistence across collaborations provides a reference point against which more variable guitar–voice interactions can be interpreted.

\section{Preliminary Musicological Insights}
\label{sec:discussion}
The degree of variation observed across the corpus is consistent with expectations for this repertoire. It is unsurprising that only a single trait—correlation between vocal melodic contour and vocal loudness—remains perceptible at a global level. Although both Carlos Paredes and the collaborating singers display recognizable timbral and expressive characteristics, each recorded song functions as a self-contained musical miniature, shaped by its own poetic, melodic, and formal specificities. Expressive traits are articulated idiosyncratically at the song level rather than uniformly across collaborations. Even within collaborations involving the same singer, recordings may diverge substantially in their structural and performative organization.

This is illustrated in the recordings with Luiz Goes. Both `Balada do Mar' and `Canção da Infância' employ ternary subdivision (12/8 and 6/8, respectively). In `Balada do Mar', Paredes constructs a chromatic pitch-descending ostinato in sixteenth notes, whereas in `Canção da Infância' he accompanies Goes with a more conventional arpeggiated pattern in eighth notes, closely aligned with the harmonic framework. While Goes’s vocal interpretation remains broadly consistent across the two recordings, Paredes adopts contrasting accompanimental approaches: the former foregrounds a continuously active, textural, and unsettled guitar gesture, while the latter reflects a more traditional and harmonically stable accompanimental role. In the analyzed descriptors, this divergence is reflected in the relationship between vocal melodic contour and guitar rhythmic density, which is positive in `Balada do Mar' and negative in `Canção da Infância', indicating contrasting configurations of melodic motion and guitar activity within the same collaboration.

At the level of individual collaborations, additional patterns emerge. In Camacho, the strongest correlations are observed within the guitars and vocal layers independently, suggesting that expressive organization is primarily articulated internally within each part rather than through persistent linear coupling. A comparable configuration is observed in Correia de Oliveira, where correlations are concentrated within the vocal layer. These results suggest that while vocal approaches remain relatively stable within collaborations, Paredes’s guitar language is not applied uniformly but selectively reconfigured in response to local structural and performative contexts. By contrast, Goes collaboration exhibits a negative correlation between voice rhythmic density and guitar loudness, pointing to a more dynamic redistribution of expressive weight between the two layers. The recordings with Melo depart further from the Coimbra-related collaborations, showing alignment between guitar and voice rhythmic density, which suggests a different mode of coordination centered on shared melodic–rhythmic intentions. Overall, these observations indicate that while singers tend to maintain a stable expressive profile within a collaboration, Paredes’s guitar adapts flexibly to the musical and stylistic demands of individual pieces. Such flexibility is consistent with his broader artistic identity as a versatile and multidisciplinary musician, whose work frequently engaged with other art forms, including theater, dance, and cinema, as well as with performers from diverse musical backgrounds \cite{Fonseca2025CarlosParedes}. 

Consequently, no single interactional signature can be attributed to the \textit{Canção de Coimbra} singers as a group—a result that is musically coherent given the stylistic diversity and limited size of the analyzed repertoire. This contrasts with score-based singer–accompanist traditions, where interactional hierarchies are largely stabilized by notation, and instead aligns with accounts of orally transmitted and improvisation-informed practices, in which musical structure and coordination emerge through performance and remain sensitive to piece-specific demands \cite{prouty2006orality, christodoulou2025amd,tani2024traditional}.

In the diversity analysis, only sparse residual outliers were detected (approximately two to three per piece). These outliers correspond to moments of substantial reorganization within individual songs, including the introduction of new harmonic changes or musical material (e.g., transitional instrumental sections, repeated ostinatos, or overlapping layers that occur only once), the most pronounced accelerandos and ritardandos relative to comparable passages, entrances of the voice, and fade-outs. Outliers were most frequently observed at the beginnings and endings of songs; when they occurred mid-piece, they consistently aligned with manually annotated formal section boundaries. This behavior is particularly evident in `Render dos Heróis' (Melo collaboration), a piece characterized by a succession of short, contrasting sections reflecting its original conception as theater music. Here, outliers capture several structurally salient events, including harmonic shifts manifested through changes in the relationship between tonal dissonance and vocal rhythmic density, the introduction of a novel instrumental passage between vocal sections, the only segment in which voice and guitar articulate the same melodic–rhythmic profile in unison (with additional harmonic elaboration in the guitar), and the final ritardando. These findings indicate that residual-based diversity measures are sensitive to moments of significant structural and performative transformation rather than to local expressive fluctuation.

\section{Conclusions and Future Work}
\label{sec:conclusions}
This paper introduced a computational framework for analyzing voice–guitar interaction in Carlos Paredes’s vocal collaborations, combining source separation, beat‑level descriptors, correlation analysis, and residual‑based diversity detection. Applied to eight recordings, the method shows that expressive organization in this repertoire is largely piece‑specific: apart from a consistent relationship between vocal melodic contour and vocal loudness, no cross‑corpus interaction pattern emerged. Diversity events identified through residual analysis aligned with musically meaningful transitions, confirming the framework’s ability to highlight structural reorganization.

Future work should expand the descriptor set—particularly with timbral features suited to Portuguese guitar—and refine model parameters to improve robustness. Extending the study to a larger or diachronic corpus may uncover broader stylistic trajectories in Paredes’s collaborations. More broadly, the framework is well‑positioned for application to repertoires where composition and performance are intertwined, offering a computational approach beyond score‑centered music analysis.


\bibliography{icmc2026_paper_template}

@inproceedings{bernardes2017automatic,
  title={Automatic musical key estimation with adaptive mode bias},
  author={Bernardes, Gilberto and Davies, Matthew EP and Guedes, Carlos},
  booktitle={IEEE ICASSP},
  pages={316--320},
  year={2017},
}

@article{debut2016physics,
  title={Physics-based modeling techniques of a twelve-string Portuguese guitar: A non-linear time-domain computational approach for the multiple-strings/bridge/soundboard coupled dynamics},
  author={Debut, Vincent and Antunes, Jos{\'e} and Marques, Mour{\~a}o and Carvalho, Mour{\~a}o},
  journal={Applied Acoustics},
  volume={108},
  pages={3--18},
  year={2016},
  publisher={Elsevier}
}

@article{2020SciPy-NMeth,
  author  = {Virtanen, Pauli and Gommers, Ralf and Oliphant, Travis E. and others},
  title   = {{SciPy} 1.0: Fundamental Algorithms for Scientific Computing in Python},
  journal = {Nature Methods},
  year    = {2020},
  volume  = {17},
  pages   = {261--272},
}

@article{cancinochacon2018comprehensive,
  title={Computational Models of Expressive Music Performance: A Comprehensive and Critical Review},
  author={Cancino-Chac{\'o}n, Carlos Eduardo and Grachten, Maarten and Goebl, Werner and Widmer, Gerhard},
  journal={Front. in Digital Humanities},
  volume={5},
  pages={25},
  year={2018},
  publisher={Frontiers Media SA},
  doi={10.3389/fdigh.2018.00025}
}

@article{palmer1997music,
  title={Music Performance},
  author={Palmer, Caroline},
  journal={Annual Review of Psychology},
  volume={48},
  number={1},
  pages={115--138},
  year={1997},
  publisher={Annual Reviews},
  doi={10.1146/annurev.psych.48.1.115}
}

@article{prouty2006orality,
  title={Orality, Literacy, and Mediating Musical Experience: Rethinking Oral Tradition in the Learning of Jazz Improvisation},
  author={Prouty, Kenneth E},
  journal={Popular Music and Society},
  volume={29},
  number={3},
  pages={317--334},
  year={2006},
  publisher={Taylor \& Francis},
  doi={10.1080/03007760600670372}
}

@book{tani2024traditional,
  title={Traditional Iranian Music: Orality, Physicality and Improvisation},
  author={Tani, Masato},
  year={2024},
  publisher={Trans Pacific Press},
  isbn={978-1920850357}
}

@article{rosenblum2023varying,
  title={The Varying Social Dynamics in Orally Transmitted and Notated vs. Improvised Musical Performance},
  author={Rosenblum, Shani and Schaffrath, Clemens and Maes, Pieter-Jan and Leman, Marc},
  journal={Front. in Psychology},
  volume={14},
  pages={1106092},
  year={2023},
  publisher={Frontiers Media SA},
  doi={10.3389/fpsyg.2023.1106092}
}

@book{meredith2016computational,
  title={Computational Music Analysis},
  author={Meredith, David},
  year={2016},
  publisher={Springer},
  address={Cham, Switzerland},
  doi={10.1007/978-3-319-25931-4}
}

@article{chen2024indepth,
  title={An In-depth Review on Music Source Separation},
  author={Chen, Yuanhao and Zhang, Shuai and Wang, Xinyi and Zhao, Tianyu and Xiang, Wenping},
  journal={arXiv:2401.11378},
  year={2024}
}

@article{schulzeforster2022unsupervised,
  title={Unsupervised Music Source Separation Using Differentiable Parametric Source Models},
  author={Schulze-Forster, Kilian and Richard, Ga{\"e}l and Doire, Clement SJ and Badeau, Roland},
  journal={IEEE/ACM Transactions on Audio, Speech, and Language Proces.},
  volume={31},
  pages={1276--1289},
  year={2022},
  publisher={IEEE},
  doi={10.1109/TASLP.2022.3221052}
}

@inproceedings{defossez2021hybrid,
  title={Hybrid Spectrogram and Waveform Source Separation},
  author={D{\'e}fossez, Alexandre},
  booktitle={Proc. of the ISMIR 2021 Workshop on Music Source Separation},
  year={2021}
}

@article{stoter2019openunmix,
  title={Open-Unmix-A Reference Implementation for Music Source Separation},
  author={St{\"o}ter, Fabian-Robert and Uhlich, Stefan and Liutkus, Antoine and Mitsufuji, Yuki},
  journal={J. of Open Source Software},
  volume={4},
  number={41},
  pages={1667},
  year={2019}
}

@article{chandna2022deeplearning,
  title={A Deep-Learning Based Framework for Source Separation, Analysis, and Synthesis of Choral Ensembles},
  author={Chandna, Pritish and Cuesta, Helena and Petermann, Darius and G{\'o}mez, Emilia},
  journal={Front. in Signal Proces.},
  volume={2},
  pages={808594},
  year={2022},
  publisher={Frontiers Media SA},
  doi={10.3389/frsip.2022.808594}
}

@inproceedings{dannenberg2002pattern,
  title={Pattern Discovery Techniques for Music Audio},
  author={Dannenberg, Roger B and Hu, Ning},
  booktitle={Proc. of the ISMIR},
  pages={63--70},
  year={2002}
}

@article{paulus2010music,
  title={Music Structure Analysis Using the Patterns},
  author={Paulus, Jouni and Klapuri, Anssi},
  journal={IEEE Transactions on Audio, Speech, and Language Proces.},
  volume={18},
  number={3},
  pages={451--462},
  year={2010},
  publisher={IEEE}
}

@inproceedings{lu2004repeating,
  title={Repeating Pattern Discovery and Structure Analysis from Acoustic Music Data},
  author={Lu, Lie and Wenyin, Liu and Zhang, Hong-Jiang},
  booktitle={Proc. of the 6th ACM SIGMM Intl. Workshop on Multimedia Information Retrieval},
  pages={275--282},
  year={2004},
  doi={10.1145/1026711.1026756}
}

@inbook{muller2015music,
    author = {M{\"u}ller, Meinard and Grosche, Peter and Jiang, Nanzhu},
    title = {Handbook of Signal Proces. in Acoustics},
    publisher = {Springer},
    year = {2015},
    chapter = {Music Structure Analysis from Acoustic Signals},
pages={305--331},
}

@article{flexer2012novelty,
  title={A Case for Predominance of Novelty-Based Segmentation for Music Structure Analysis},
  author={Flexer, Arthur and Pampalk, Elias and Widmer, Gerhard},
  journal={Music Information Retrieval Exchange (MIREX)},
  year={2012}
}

@inproceedings{dittmar2012audio,
  title={Audio-Based Drum Transcription using Dynamical Models of Percussive Sounds},
  author={Dittmar, Christian and Gartner, Daniel},
  booktitle={Proc. of the Sound and Music Computing Conf.},
  year={2012}
}

@book{amiot2016music,
  title={Music Through Fourier Space: Discrete Fourier Transform in Music Theory},
  author={Amiot, Emmanuel},
  year={2016},
  publisher={Springer},
  address={Cham, Switzerland},
  series={Computational Music Science},
  doi={10.1007/978-3-319-45581-5}
}

@inproceedings{inacio2004vibroacoustic,
  title={The Portuguese Guitar Acoustics: Part 1--Vibroacoustic Measurements},
  author={In{\'a}cio, Ot{\'\i}lio and Santiago, F and Cabral, PC},
  booktitle={Proc. of the IV Iberoamerican Acoustics Congress},
  year={2004}
}

@article{serraosilva2017experimental,
  title={Experimental-numerical correlation of the dynamic behavior of the Portuguese guitar},
  author={Serr{\~a}o, Pedro and Silva, A and Infante, V and Ribeiro, AMR},
  journal={Applied Acoustics},
  volume={116},
  pages={136--148},
  year={2017},
  publisher={Elsevier}
}

@article{bernardes2016multi,
  title={A multi-level tonal interval space for modelling pitch relatedness and musical consonance},
  author={Bernardes, Gilberto and Cocharro, Diogo and Caetano, Marcelo and Guedes, Carlos and Davies, Matthew EP},
  journal={J. of New Music Research},
  volume={45},
  number={4},
  pages={281--294},
  year={2016},
  publisher={Taylor \& Francis}
}

@inproceedings{ramires2020tiv,
  title={TIV. LIB: An open-source library for the tonal description of musical audio},
  author={Ramires, Ant{\'o}nio and Bernardes, Gilberto and Davies, Matthew EP and Serra, Xavier},
  booktitle={Proc. of the Digital Audio Effects Conf.},
pages={304-309},
  year={2020}
}

@inproceedings{mauch2014pyin,
  title={pYIN: A fundamental frequency estimator using probabilistic threshold distributions},
  author={Mauch, Matthias and Dixon, Simon},
  booktitle={IEEE ICASSP},
  pages={659--663},
  year={2014},
}

@inproceedings{fitzgerald2010harmonic,
  title={Harmonic/percussive separation using median filtering},
  author={Fitzgerald, Derry},
  year={2010},
  booktitle={Proc. of the Digital Audio Effects Conf.},
  pages={1-4},
  volume={13}
}

@inproceedings{rouard2023hybrid,
  title={Hybrid transformers for music source separation},
  author={Rouard, Simon and Massa, Francisco and D{\'e}fossez, Alexandre},
  booktitle={IEEE ICASSP},
  pages={1--5},
  year={2023},
  organization={IEEE}
}

@article{mcfee2015librosa,
  title={librosa: Audio and music signal analysis in python.},
  author={McFee, Brian and others},
  journal={SciPy},
  volume={2015},
  pages={18--24},
  year={2015}
}

@book{Correia2016CancaoCoimbra,
  author    = {Correia, Andr{\'e} R.},
  title     = {Do Choupal at{\'e} {\`a} Lapa: Para uma Etnografia do Constructo da Can{\c{c}}{\~a}o de Coimbra},
  publisher = {Edi{\c{c}}{\~o}es S{\'\i}labo},
  year      = {2016},
  address   = {Lisboa},
  language  = {portuguese}
}

@book{Fonseca2025CarlosParedes,
  author    = {Fonseca, Oct{\'a}vio},
  title     = {Carlos Paredes: A Guitarra de um Povo},
  publisher = {Tradisom},
  year      = {2025},
  address   = {Portugal},
  language  = {portuguese}
}

@InProceedings{SonicVisualiser,
  author =    {C. Cannam and C. Landone and M. Sandler},
  title =     {Sonic Visualiser: An Open Source Application for Viewing, Analysing, and Annotating Music Audio Files},
  pages =     {1467--1468},
  booktitle = {Proc. of the ACM Multimedia 2010 Intl. Conf.},
    year =      {2010},
}

@misc{Paredes2002Integral,
  author       = {{C. Paredes}},
  title        = {O Mundo Segundo Carlos Paredes: Integral 1958--1993},
  howpublished = {8-CD compilation},
  publisher    = {EMI},
  year         = {2002},
  address      = {Portugal},
  note         = {Catalogue number: EMI 5 80304 2}
}

@article{christodoulou2024multimodal,
  title={Multimodal music datasets? Challenges and future goals in music processing},
  author={Christodoulou, Anna-Maria and Lartillot, Olivier and Jensenius, Alexander R.},
  journal={Intl. J. of Multimedia Information Retrieval},
  volume={13},
  year={2024}
}

@inproceedings{christodoulou2025amd,
  title={A multimodal dataset of Greek folk music},
  author={Christodoulou, Anna-Maria and Lartillot, Olivier},
  booktitle = {Proceedings of the 12th Intl. Conf. on Digital Libraries for Musicology},
pages = {19--27},
  year={2025}
}

@article{MouraVidal2023,
  author    = {Moura, N. and Vidal, M. and Aguilera, A. M. and Vilas-Boas, J. P. and Serra, S. and Leman, M.},
  title     = {Knee flexion of saxophone players anticipates tonal context of music},
  journal   = {npj Science of Learning},
  volume    = {8},
  number    = {1},
  year      = {2023},
  doi       = {10.1038/s41539-023-00172-z}
}

\end{document}